\documentclass[a4paper,11pt]{article}
\usepackage{pos}

\usepackage{amsmath}

\title{Anomalies in Particle Physics}

\author*[a,b]{Andreas Crivellin}

\affiliation[a]{Physik-Institut, Universität Zürich, \\Winterthurerstrasse 190, CH–8057 Zürich, Switzerland}

\affiliation[b]{Paul Scherrer Institut,\\ CH–5232 Villigen PSI, Switzerland}

\emailAdd{andreas.crivellin@cern.ch}

\abstract{I provide a (personal) review of the current hints for physics beyond the Standard Model, called ``anomalies'', obtained both at the intensity frontier (flavour and electroweak precision observables) and in direct LHC searches. This includes the deviations from the Standard Model predictions in semi-leptonic $B$ decays, the anomalous magnetic moment of the muon, the Cabibbo Angle Anomaly, the $W$ mass as well as non-resonant di-lepton searches, the hints for new scalar particles around $\approx\! 95\,$GeV, $\approx\! 151\,$GeV, $\approx\! 670\,$GeV and the (di-)di-jet excess at $\approx \!1\,$TeV ($\approx 3.6\,$TeV). Possible explanations in terms of new particles are briefly summarized and discussed. }

\FullConference{
8${}^{\rm th}$ Symposium on Prospects in the Physics of Discrete Symmetries (DISCRETE 2022)\\
  7-11 November 2022\\ Baden-Baden, Germany}

\begin{document}
\maketitle

The Standard Model (SM) of particle physics describes the known fundamental constituents of matter as well as their interactions (excluding gravity). It has been extensively tested and verified by experiments within the last decades~\cite{ParticleDataGroup:2022pth}, with the discovery of the Higgs boson at the Large Hadron Collider (LHC) at CERN in 2012 being the last missing piece of the puzzle~\cite{ATLAS:2012yve,CMS:2012qbp}. Even though the SM is now complete, it cannot be the ultimate fundamental theory of Nature: In addition to theoretical arguments for the existence of beyond the SM (BSM) physics, the SM e.g.~cannot account for the observations of Dark Matter (DM) established at cosmological scales, nor for the non-vanishing neutrino masses required by neutrino oscillations.

While so far no such new states have been unequivocally observed, in recent years an increasing number of hints for BSM physics, called “anomalies”, have been reported. They span over a huge energy range, from precision measurements of muon properties (the anomalous magnetic moment of the muon), over semi-leptonic $B$ meson decays, the measurement of the $W$ boson mass, to direct LHC searches. While probability theory and statistics tell us that one cannot expect that all these anomalies will be confirmed in the future, it is also unlikely that all of them are just statistical flukes. Therefore, it is important to assess to which extensions of the SM these anomalies point to, for progressing towards a unified explanation that could also e.g.~account for neutrino mass, DM etc. In these proceedings, I will review the status of these anomalies and give an overview of how they can be explained by extending the SM with new particles and new interactions. 

\section{Anomalies}

Let us now review the status of these anomalies. We will present them in increasing order of the corresponding energy scale.

\subsection{Anomalous magnetic moment of the muon ($a_\mu$)}

The combined value of the Brookhaven E821 result~\cite{Bennett:2006fi} and the $g-2$ experiment at Fermilab~\cite{Abi:2021gix} for the anomalous magnetic moment of the muon ($a_\mu$) shows a $4.2\,\sigma$ tension with the SM prediction~\cite{Aoyama:2019ryr}. However, this SM prediction of the $g-2$ theory initiative is based on $e^+e^-\to$hadrons~\cite{Colangelo:2018mtw,Davier:2019can,Keshavarzi:2019abf} and does not include later lattice QCD results for hadronic vacuum polarization~\cite{Borsanyi:2020mff} nor the last measurement of $e^+e^-\to$hadrons~\cite{CMD-3:2023alj} which would render the SM prediction closer to the measurement. 

Since the limit on NP from $a_e$ is so stringent~\cite{Hanneke:2008tm,Aoyama:2017uqe,Laporta:2017okg}, the effect in $a_\ell$ cannot be flavour blind (like the Schwinger term). Therefore, to explain $a_\mu$, NP must violate lepton flavour universality (LFU), however, the breaking of LFU could originate from the SM Yukawa couplings, like it is for example the case in the MSSM with minimal flavour violation.

\subsection{Cabibbo Angle Anomaly (CAA)}

The Cabibbo angle parametrizes the mixing between the first two quark generations and in particular dominates the first-row and first-column CKM unitarity relations. These relations can be used to check the consistency of different determinations of CKM elements that are sensitive to NP contributions. Interestingly, in addition to a deficit in first-row and first-column CKM unitarity, which can be traced back to the fact that $V_{ud}$ extracted from super-allowed beta decays does not agree with $V_{us}$ determined from kaon and tau decays, (when comparing them via CKM unitarity), there is also a disagreement between the determinations of $V_{us}$ from $K_{\ell2}$ and $K_{\ell3}$, as can be seen in the left plot of Fig.~\ref{bsll}~\cite{Cirigliano:2022yyo}.\footnote{The significance of these deviations crucially depends on the radiative corrections to beta decays~\cite{Marciano:2005ec,Seng:2018yzq,Hardy:2020qwl} and on the treatment of the tensions between $K_{\ell 2}$ and $K_{\ell 3}$~\cite{Moulson:2017ive,Seng:2021nar} and tau decays~\cite{Amhis:2019ckw}.}

While the former can be explained via left-handed new physics in beta decays, the latter requires a right-handed $\bar d s$ current. Alternatively, the unitarity deficit can be interpreted as a sign of LFU violation~\cite{Coutinho:2019aiy,Crivellin:2020lzu}, because beta decays involve electrons while the best determination of $V_{us}$ comes from $K\to\mu\nu$. Furthermore, as for the extraction of $V_{ud}$ from beta decays knowledge of the Fermi constant is necessary, an interplay with the global EW fit, where this parameter is one of the crucial inputs, arises~\cite{Crivellin:2021njn}.

\subsection{$\tau\to\mu\nu\nu$}

Combining the ratios of ${\rm Br}[\tau  \to \mu(e) \nu \bar \nu]/{\rm Br}[\mu \to e \nu \bar \nu]$ and ${\rm Br}[\tau  \to \mu \nu \bar \nu]/{\rm Br}[\tau \to e \nu \bar \nu]$~\cite{Belle:2013teo,Amhis:2019ckw}, leads to a $\approx 2\sigma$ preference for constructive NP at the per-mille level in $\tau  \to \mu \nu \bar \nu$~\cite{Bryman:2021teu}.

\subsection{$ b\!\!\to\!\! c\ell\nu$}

These charged current transitions, mediated at tree-level in the SM, have significant branching ratios (${\mathcal O}(10^{-3})$). Here, the ratios $R\left( {{D^{\left( * \right)}}} \right)$ = ${\rm Br}({B \to {D^{\left( * \right)}}\tau \nu })$/ ${\rm Br}({ B \to {D^{\left( * \right)}}\ell \nu )})$, are bigger than the SM predictions by approximately 20\%, resulting in a $\gtrapprox3\sigma$ significance~\cite{HFLAV:2022pwe}.

\subsection{$b\to s\ell^+\ell^-$}

Like all flavour changing neutral current processes, $b\to s\ell^+\ell^-$ transitions are loop and CKM suppressed within the SM, resulting in small branching ratios, up to a few times $10^{-6}$. While the previous hints~\cite{LHCb:2021lvy} for lepton flavour universality violating in the ratios 
$R({K^{\left(*\right)}})  = {\rm Br}({B \to K^{\left( * \right)}\mu ^+\mu ^-})
/{\rm Br}({B \to K^{\left( * \right)}e^+e^-})$
were not confirmed~\cite{LHCb:2022qnv} and $B_s\to\mu^+\mu^-$~\cite{LHCb:2020zud,CMS:2022mgd} now agrees quite well with the SM prediction~\cite{Hermann:2013kca,Beneke:2017vpq}, there are several LFU $b \to s \ell^+\ell^-$ observables that significantly deviate from the SM predictions, most importantly the angular observable $P_5^{\prime}$~\cite{Descotes-Genon:2012isb,LHCb:2020lmf}, the total branching ratio ${\rm Br}({B \to K\mu^+\mu^-})$\cite{LHCb:2014cxe,Parrott:2022zte} and ${\rm Br}({B_s \to \phi\mu^+\mu^-})$~\cite{LHCb:2021zwz,Gubernari:2022hxn}. 

\begin{figure}
	\centering
	\vspace{-3mm}
    \includegraphics[width=0.5\textwidth]{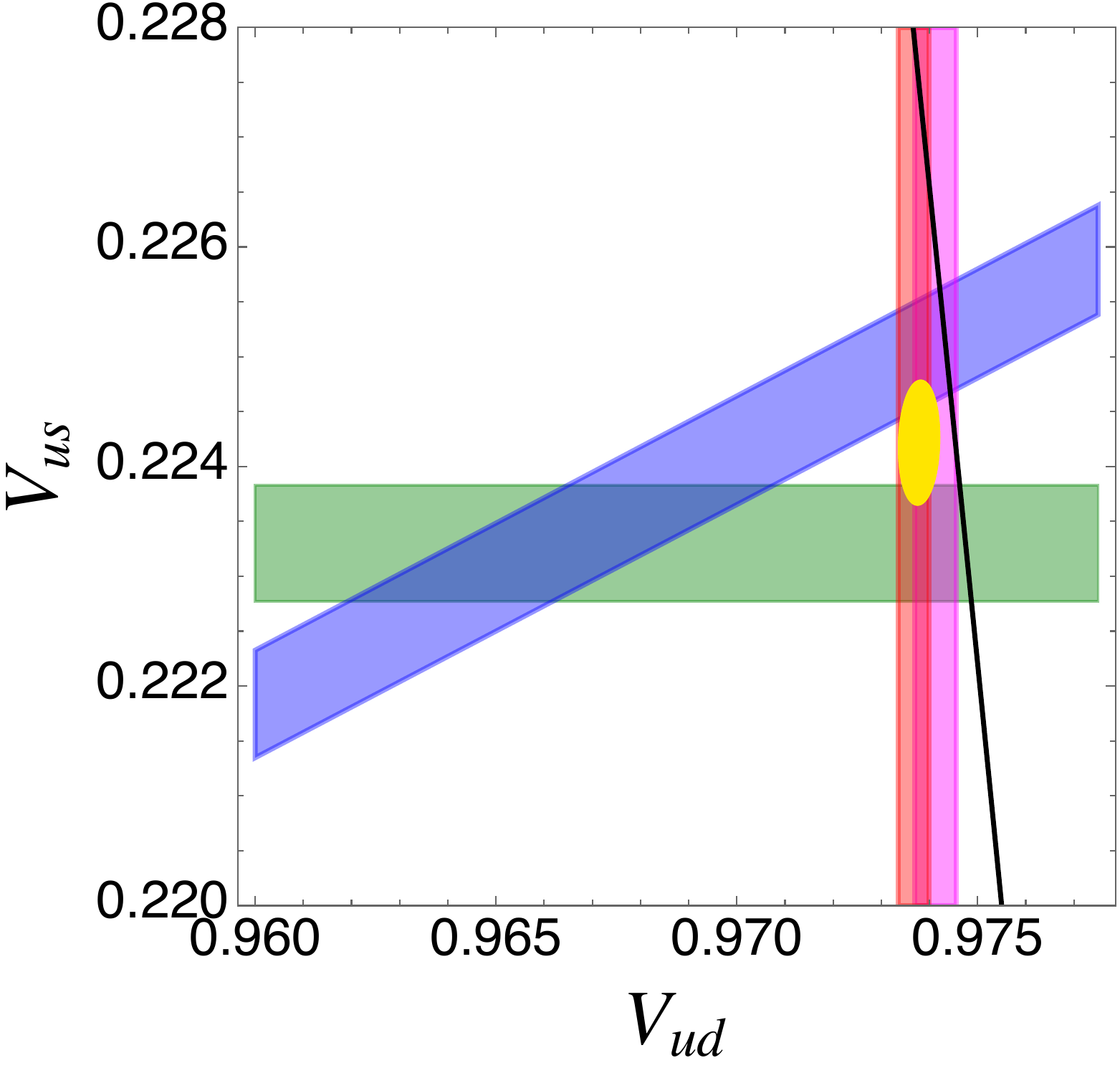}
	\includegraphics[width=0.48\textwidth]{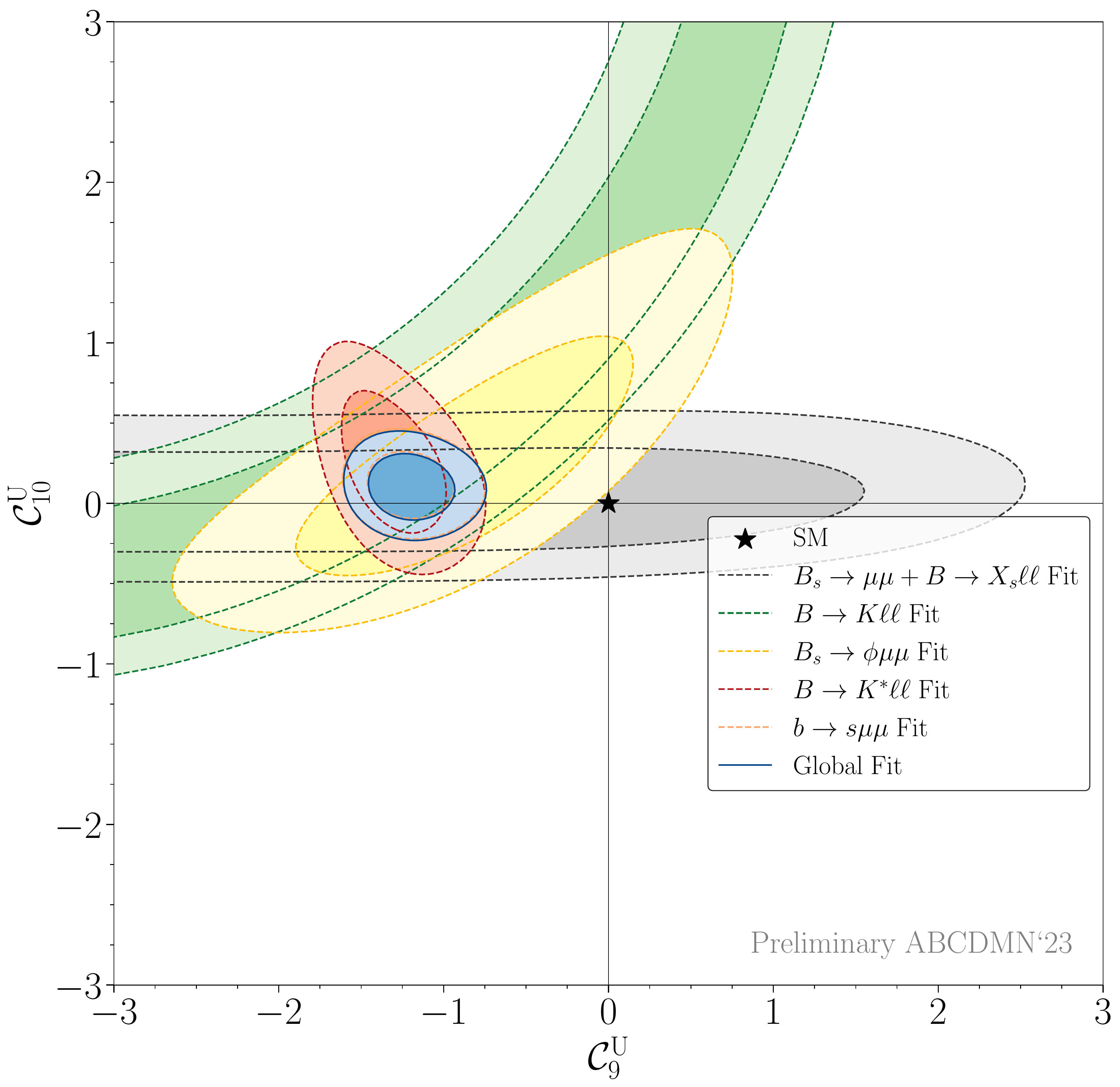}
	\vspace{-3mm}
	\caption{Left: Constraints in the $V_{ud}$--$V_{us}$ plane. The partially overlapping vertical bands correspond to $V_{ud}^{0^+\to 0^+}$ (leftmost, red) and $V_{ud}^\text{n, best}$ (rightmost, violet). 	The horizontal band (green) corresponds to $V_{us}^{K_{\ell3}}$. 	The diagonal band (blue) corresponds to $({V_{us}}/{V_{ud}})_{K_{\ell 2}/\pi_{\ell 2}}$. 	The unitarity circle is denoted by the black solid line. 	The $68\%$ C.L.\ ellipse, from a fit to all four constraints, is depicted in yellow and deviates from the unitarity line by $2.8\sigma$. Note that the significance tends to increase in case $\tau$ decays are included.  Right: Global fit to $b\to s\ell^+\ell^-$ data showing that the scenario $C_9^{\rm U}$ describes data much better than the SM hypothesis.~\cite{QuimPreliminary}
	}
	\label{bsll}
\end{figure}

While $R({K^{\left(*\right)}})$ point towards LFU NP and $B_s\to \mu\mu$ to no NP in $C_{10}$, the other observables can be explained by $C_{9}$, i.e.~by a left-handed $\bar sb$ and a vectorial lepton current. In fact, including all observables into a global fit (see Fig.~\ref{bsll}), one finds that the simple scenario with a LFU contribution to $C_9$, i.e.~$C_9^{U}$, is preferred over the SM hypothesis by more than $6\,\sigma$~\cite{Alguero:2019ptt,Buras:2022qip,Ciuchini:2022wbq} (see the right plot in Fig.~\ref{bsll}).

\subsection{$Z\to b \bar b$}

The LEP measurement of the forward-backward asymmetry in $Z\to b\bar b$~\cite{ALEPH:2005ab} deviates from its SM prediction by $\approx2\sigma$. Similarly, there is a $\approx 2\sigma$ tension in the lepton asymmetry parameter $A_\ell$~\cite{deBlas:2022hdk}, mainly due to the electron channel ($A_e$).

\subsection{$W$ boson mass ($m_W$)}

Combining the new CDF result~\cite{CDF:2022hxs} with the previous ones from the LHC~\cite{ATLAS:2017rzl,CMS:2011utm,LHCb:2015jyu,LHCb:2021bjt}, one finds $m_W=(80.413\pm 0.015)$GeV, employing a conservative error estimate, which corresponds to a $3.7\,\sigma$ tension with the SM prediction~\cite{deBlas:2022hdk}.\footnote{This average does not include the latest ATLAS result (ATLAS-CONF-2023-004).}

\subsection{Neutral scalars at the LHC ($\gamma\gamma$)}

There are several hints for new resonances at the LHC, mainly in di-photon channels at $95\,$GeV~\cite{CMS:2023yay}, 151$\,$GeV~\cite{ATLAS:2021jbf,Crivellin:2021ubm} and $\approx 670\,$GeV~\cite{ATLAS:2021uiz}. The hint at $95\,GeV$ is supported by an di-taus excess at CMS~\cite{CMS:2022rbd}, an $ZH$ one (with $H\to b\bar b$) by LEP~\cite{LEPWorkingGroupforHiggsbosonsearches:2003ing} as well as $WW$~\cite{Coloretti:2023wng}. The hint for a new scalar at $\approx 670\,GeV$ is also supported by a $ZZ\to4\ell$ measurement~\cite{CMS:2017dib} and the CMS di-Higgs excess at $\approx680\,$GeV in 
 an asymmetric $\gamma\gamma b\bar b$ search~\cite{CMS:2022tgk}. Here, the best fit is obtained for an invariant $b\bar b$ mass of $90\,$GeV, i.e.~compatible with the $95\,$GeV $\gamma\gamma$ excess, taking into account the bottom jet energy resolution.

\subsection{(Di-)di-jets (jj(jj))}

ATLAS~\cite{ATLAS:2018qto} observed a weaker limit than expected in resonant di-jet searches\footnote{The analogous CMS di-jet search~\cite{CMS:2018mgb} does not display an excess in the same region. However, the sensitivity is significantly lower, such that the signal suggested by the ATLAS analysis is not excluded.} in a mass region slightly below $1\,$TeV, while CMS~\cite{CMS:2022usq} found hints for the (non-resonant) pair production of di-jet resonances with a mass of $\approx 950$\,GeV with a local (global) significance of 3.6\,$\sigma$ (2.5\,$\sigma$) when integrating over the di-di-jet mass.

While the ATLAS analysis alone does not constitute a significant hint for BSM physics once the look-elsewhere effect (LEE) is taken into account, the compatibility of the suggested di-jet mass with the one of the (non-resonant) CMS di-di-jet analysis is very good. This agreement suggests that both excesses might be due to the same new particle $X$, once directly (resonantly) produced in proton-proton collisions ($pp\!\to\! X\!\to\! jj$), once pair produced via a new state $Y$ ($pp\!\to\! Y^{(*)}\!\to\! XX\!\to\! (jj)(jj)$). While the CMS collaboration in their analysis interprets the di-di-jet excess as the non-resonant production of $XX$ (with $m_X\approx 950\,$GeV) via a heavy new particle $Y$, with $m_Y\approx 8$\,TeV, resulting in a local (global) significance of $3.9\sigma$ ($1.6\sigma$)~\cite{CMS:2022usq}, it is also possible that the two $X$ particles are resonantly produced from the decay of an on-shell $Y$ particle. In fact, the CMS results suggest 3\,TeV$\lessapprox \!m_Y\!\lessapprox$4\,TeV for such a resonant scenario, once $m_X$ is assumed to be within the preferred range of the ATLAS di-jet analysis and  Ref.~\cite{Crivellin:2022nms} found a $4\,\sigma$ significance at $\approx 3.6\,$TeV.

\subsection{$q\bar q\to e^+e^-$}

In the non-resonant search for high-energetic oppositely charged lepton pairs, CMS observed 44 electron events with an invariant mass above $1.8\,$TeV, while only $29.2\pm3.6$ events were expected~\cite{CMS:2021ctt}. As the number of observed muons is compatible with the SM prediction, this is another sign of LFUV and CMS provided the ratio of muons over electrons which reduces the theoretical uncertainties~\cite{Greljo:2017vvb}. Importantly, this CMS excess is compatible with the corresponding ATLAS limit~\cite{ATLAS:2020yat} where also slightly more electrons than expected were observed. Performing a model-independent fit, one finds the NP at a scale of $10\,$TeV with order one couplings can improve over the SM hypothesis by $\approx3\,\sigma$~\cite{Crivellin:2021rbf}.

\subsection{Summary}

\begin{figure}
	\centering
	\vspace{-3mm}
	\includegraphics[width=0.96\textwidth]{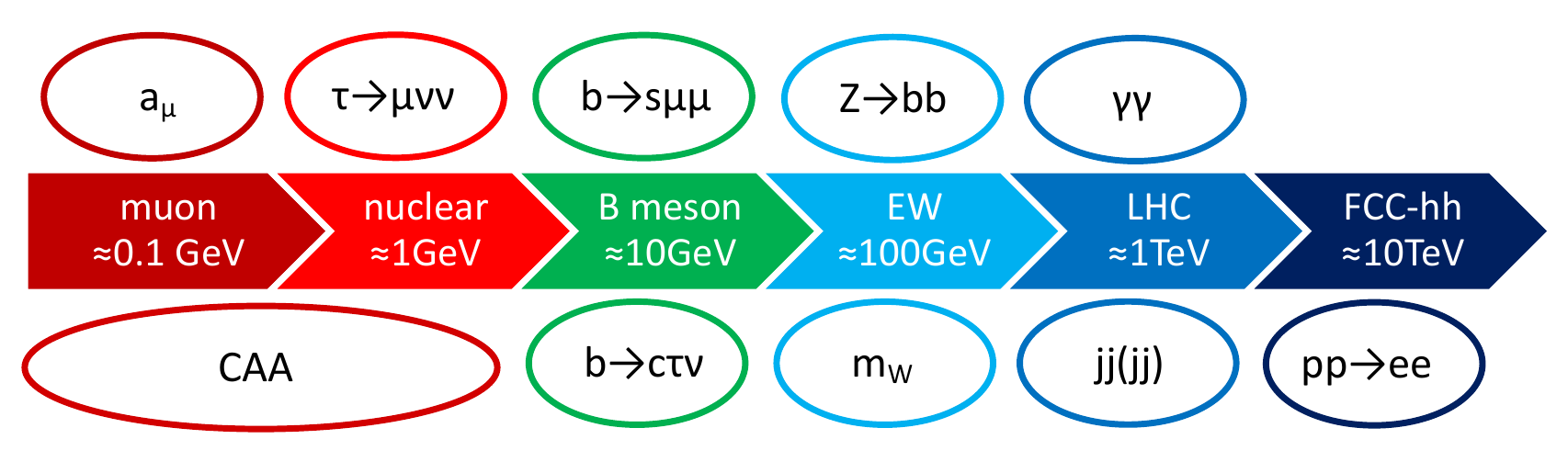}
	\vspace{-3mm}
	\caption{Compilation of various anomalies ordered according to the corresponding energy scale. 
	}
	\label{anomalies}
\end{figure}

The anomalies are summarized in Fig.~\ref{anomalies}, together with their corresponding energy scale, showing that they range at least 5 orders of magnitude.
	
\section{New Physics Explanations}

For a consistent renormalizable extension of the SM, only scalars bosons (spin 0), fermions (spin 1/2) and vectors bosons (spin 1) are at one's disposal, provided that in the latter case a Higgs-like mechanism of spontaneous symmetry-breaking exists. In these proceedings we focus on five classes of SM extensions in terms of heavy NP (i.e.~realized above the EW breaking scale):

\begin{itemize}
	\item Leptoquarks ({\bf LQs}): Scalar or vector particles that carry color and couple quarks directly to leptons~\cite{Buchmuller:1986zs,Dorsner:2016wpm}. These particles were first proposed in the context of the Pati-Salam model~\cite{Pati:1974yy}, Grand Unified Theories (GUTs)~\cite{Georgi:1974sy} and in the R-parity violating MSSM (see e.g.~Ref.~\cite{Barbier:2004ez} for a review).
	\item $W^\prime$ bosons: Electrically charged but QCD neutral vector particles. They appear e.g.~as Kaluza-Klein excitations of the SM $W$ in composite~\cite{Weinberg:1962hj} or extra-dimensional models~\cite{Randall:1999ee} but also in models with additional $SU(2)_L$ factors, including left-right-symmetric models~\cite{Mohapatra:1974gc}.
	\item $Z^\prime$ bosons: Neutral heavy vector bosons. They can be singlets under $SU(2)_L$ but also the neutral component of an $SU(2)_L$ multiplet. Again, these particles can be resonances of the SM $Z$ or originate from an abelian symmetry like $B-L$~\cite{Pati:1974yy} or gauged flavour symmetries~\cite{Froggatt:1978nt}.
	\item Vector-like fermions ({\bf VLF}): For vector-like fermions left-handed and right-handed fields have the same quantum numbers under the SM gauge group and can thus have masses independently of the EW symmetry breaking. They appear in GUTs~\cite{Langacker:1980js}, as resonances of SM fermions in composite or extra-dimensional models~\cite{Antoniadis:1990ew} and as the supersymmetric partners of SM vectors and scalars~\cite{Haber:1984rc}. Furthermore, vector-like leptons are involved in the type I~\cite{Minkowski:1977sc,Lee:1977tib} and type III~\cite{Foot:1988aq} seesaw mechanisms. 
	\item New scalars ($S$): Scalars could be supersymmetric partners of SM fermions~\cite{Haber:1984rc}, but also scalar fields of different representations can be added to the SM, most commonly an $SU(2)_L$ doublet leading to a 2HDM~\cite{Chanowitz:1985ug,Branco:2011iw}. Note that we do not include scalars with the properties of LQs here.
\end{itemize}

\subsection{$a_\mu$}

Because the deviation from the SM prediction is as large as its EW contribution, heavy NP at TeV scale must possess an enhancement factor (see Ref.~\cite{Athron:2021iuf} for a recent overview on NP in $a_\mu$). This can be provided via the mechanism of chiral enhancement, meaning that the chirality flip does not originate from the small muon Yukawa coupling but rather from a larger coupling of other particles to the SM Higgs. In the MSSM, this factor is $\tan\beta$~\cite{Everett:2001tq,Feng:2001tr}, while also models with generic new scalars and fermions can explain $a_\mu$~\cite{Czarnecki:2001pv,Kannike:2011ng,Kowalska:2017iqv,Crivellin:2018qmi,Crivellin:2021rbq}. Furthermore, and there are two scalar LQs ($S_1$ and $S_2$) that address $a_\mu$ via a $m_t/m_\mu$ enhancement~\cite{Djouadi:1989md,Bauer:2015knc,ColuccioLeskow:2016dox,Crivellin:2020tsz}.

\subsection{CAA}

A sub per-mille effect affecting the determination of $V_{ud}$ suffices to explain the CAA. Because in order to determine $V_{ud}$ from beta decays knowledge of the Fermi constant, most precisely measured in muon decay~\cite{Tishchenko:2012ie}, is needed, we have the following possibilities~\cite{Crivellin:2021njn}: 1) a direct (tree-level) modification of beta decays 2) a direct (tree-level) modification of muon decay 3) a modified $W$-$\mu$-$\nu$ coupling 4) a modified $W$-$u$-$d$ coupling. Option 1) could in principle be realized by a $W^\prime$~\cite{Capdevila:2020rrl} or a LQ~\cite{Crivellin:2021egp}, however in the latter case stringent bounds from other flavour observables arise. Possibility 2) can be achieved by adding a singly charged $SU(2)_L$ singlet scalar~\cite{Crivellin:2020klg}, a $W^\prime$~\cite{Capdevila:2020rrl} or $Z^\prime$ boson with flavour violating couplings~\cite{Buras:2021btx}, while option 3) and 4) can be achieved by vector-like leptons~\cite{Crivellin:2020ebi,Kirk:2020wdk} and vector-like quarks~\cite{Belfatto:2019swo,Branco:2021vhs,Belfatto:2021jhf,Crivellin:2022rhw}, respectively. Note that vector-like quarks could also solve the tension between the different determinations of $V_{us}$ while vector-like leptons have the potential to improve the global EW fit.

\subsection{$\tau\to\mu\nu\nu$}~Explanations of $\tau\to\mu\nu\nu$ are very similar to the ones of the CAA via a modified Fermi constant (with $\tau\to\mu\nu\nu$ taking the role of $\mu\to e\nu\nu$). It can therefore be achieved by a tree-level effect via a singly charged $SU(2)_L$ singlet scalar~\cite{Crivellin:2020klg}, a $W^\prime$ or a flavour violating $Z^\prime$~\cite{Buras:2021btx}. Alternatively, a modification of the $W$-$\tau$-$\nu$ coupling  through the mixing of vector-like leptons or a $W^\prime$ boson is possible. Furthermore, a $Z^\prime$ boson coupling to muons and tau leptons can generate the desired effect via box diagrams~\cite{Altmannshofer:2014cfa}.

\subsection{$b\to c\ell\nu$}

Because this transition occurs at tree-level in the SM, also a tree-level NP effect is necessary to obtain the needed effect of $O(10)$\% w.r.t.~the SM (assuming heavy NP with perturbative couplings). As this is a charged current process, the only options are charged Higgses~\cite{Crivellin:2012ye,Fajfer:2012jt,Celis:2012dk}, $W’$~bosons~\cite{Bhattacharya:2014wla} (with or without right-handed neutrinos) or LQs~\cite{Sakaki:2013bfa,Bauer:2015knc,Freytsis:2015qca,Fajfer:2015ycq}. The first two possibilities are disfavoured by the $B_c$ lifetime~\cite{Celis:2016azn,Alonso:2016oyd} and/or LHC searches~\cite{Bhattacharya:2014wla,Greljo:2015mma} (while a partial explanation is still possible~\cite{Iguro:2022uzz,Blanke:2022pjy}), leaving LQ as the best solutions. However, also for LQs constraints from $B_s-\bar B_s$ mixing, $B\to K^{(*)}\nu\nu$ and LHC searches must be respected. Therefore, the $SU(2)_L$ singlet vector LQ~\cite{Calibbi:2015kma,Barbieri:2016las,DiLuzio:2017vat,Calibbi:2017qbu,Bordone:2017bld,Blanke:2018sro,King:2021jeo} or the singlet-triplet model~\cite{Crivellin:2017zlb,Crivellin:2019dwb,Gherardi:2020qhc} are particularly interesting. 

An explanation of $\Delta A_{FB}$ needs a non-zero Wilson coefficient of the tensor operators. Importantly, among renormalizable models, only two scalar LQ can generate the corresponding dimension-6 operator at tree-level and only the $SU(2)_L$ singlet $S_1$ gives a good fit to data~\cite{Carvunis:2021dss}. However, even in this case, due to the constraints from other asymmetries, $\Delta A_{FB}$ cannot be fully explained, but the global fit to $b\to c\mu\nu$ and $b\to c e\nu$ data can be improved by more than $3\sigma$~\cite{Carvunis:2021dss}.

\subsection{$b\to s\ell^+\ell^-$}

The required $O(20\%)$ LFU NP effect w.r.t.~the SM in $C_{9}$ can be most naturally obtained via~\cite{Alguero:2022wkd}:

1)~A $Z’$ boson with LFU couplings but flavour violating couplings to bottom and strange quarks~\cite{Buras:2013qja,Gauld:2013qba}. In fact, even though one, in general, expects an effect in $B_s-\bar B_s$ mixing~\cite{DiLuzio:2017fdq}, and the $Z^\prime$ can be produced resonantly at the LHC (see e.g.~\cite{Allanach:2015gkd}), such a solution is viable if the couplings to first generations quarks are suppressed and the models possesses an approximate global $U(2)$ flavour symmetry to protect it from dangerously larger effects in $K^0-\bar K^0$ and $D^0-\bar D^0$ mixing~\cite{Calibbi:2019lvs}. 

2)~$\tau$-loop effect via off-shell photon effect in $C_9^U$~\cite{Bobeth:2014rda}. The LQ representations which can achieve this are $S_2$~\cite{Crivellin:2022mff}, $U_1$~\cite{Crivellin:2018yvo} and $S_1+S_3$~\cite{Crivellin:2019dwb}. Alternatively, in the 2HDM with generic flavour structure~\cite{Crivellin:2013wna}, a charm loop~\cite{Jager:2017gal} can generate $C_9^U$~\cite{Iguro:2023jju}.

\subsection{$Z\to b \bar b$}

An explanation of $Z\to b\bar b$ requires vector-like quarks mixing with the SM ones at tree-level as the necessary NP effect is sizable~\cite{Choudhury:2001hs,Batell:2012ca,Crivellin:2020oup}. $A_e$ could, in addition to vector-like leptons, be explained by $Z-Z^\prime$ mixing.

\subsection{$m_W$}

The tension in the $W$ mass is most easily explained by tree-level effect, e.g.~an $SU(2)_L$ scalar triplet that acquires a vacuum expectation value~\cite{Strumia:2022qkt} or via $Z-Z^\prime$ mixing in case $Z^\prime$ is an $SU(2)_L$ singlet~\cite{Alguero:2022est}. However, also loop effects of new particles with masses below or at the TeV scale are possible~\cite{Strumia:2022qkt}, such as vector-like quarks~\cite{Crivellin:2022fdf} or leptoquarks~\cite{Crivellin:2020ukd}.

\begin{figure}
    \centering    
    \includegraphics[scale=0.5]{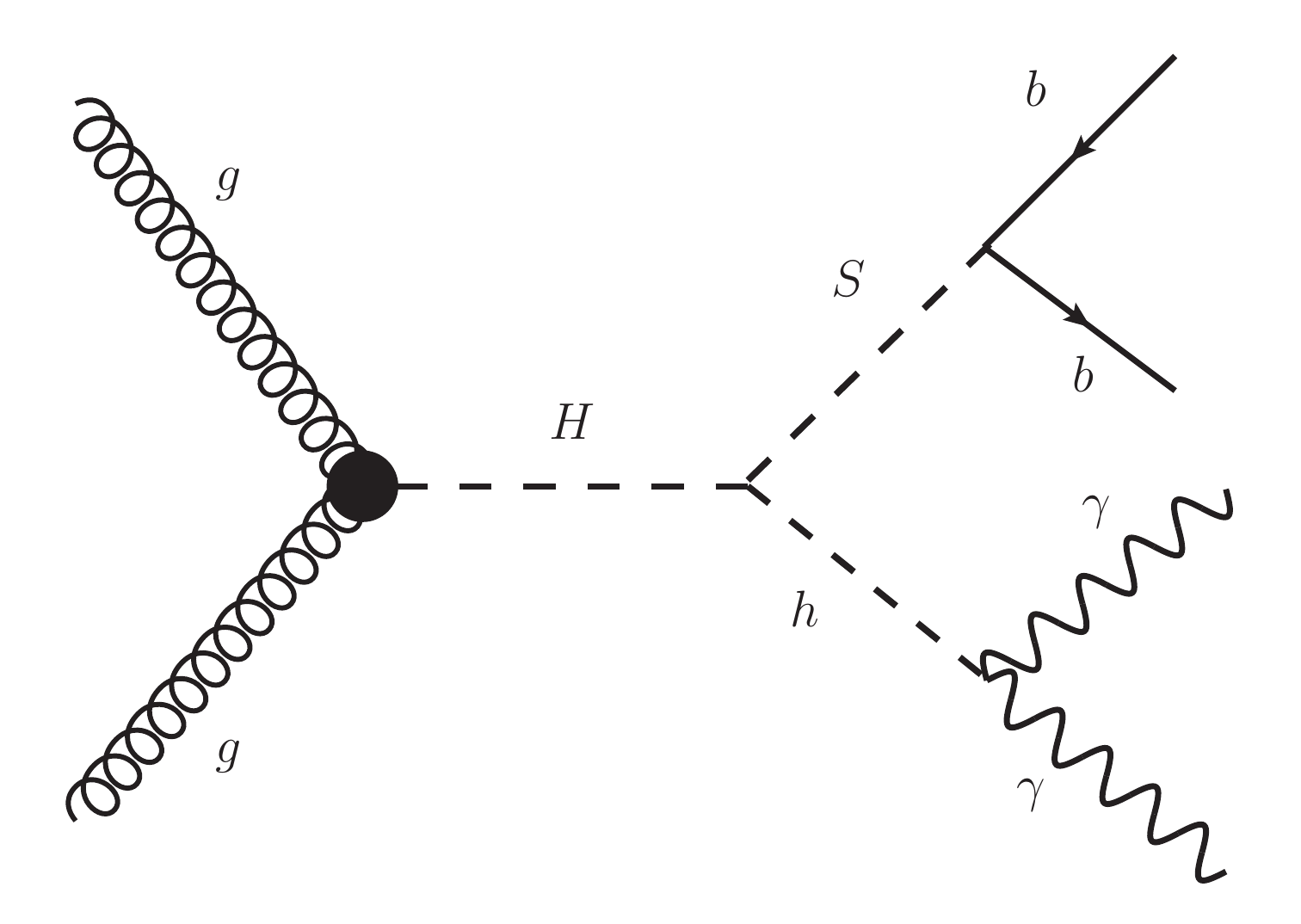}
     \includegraphics[scale=0.4]{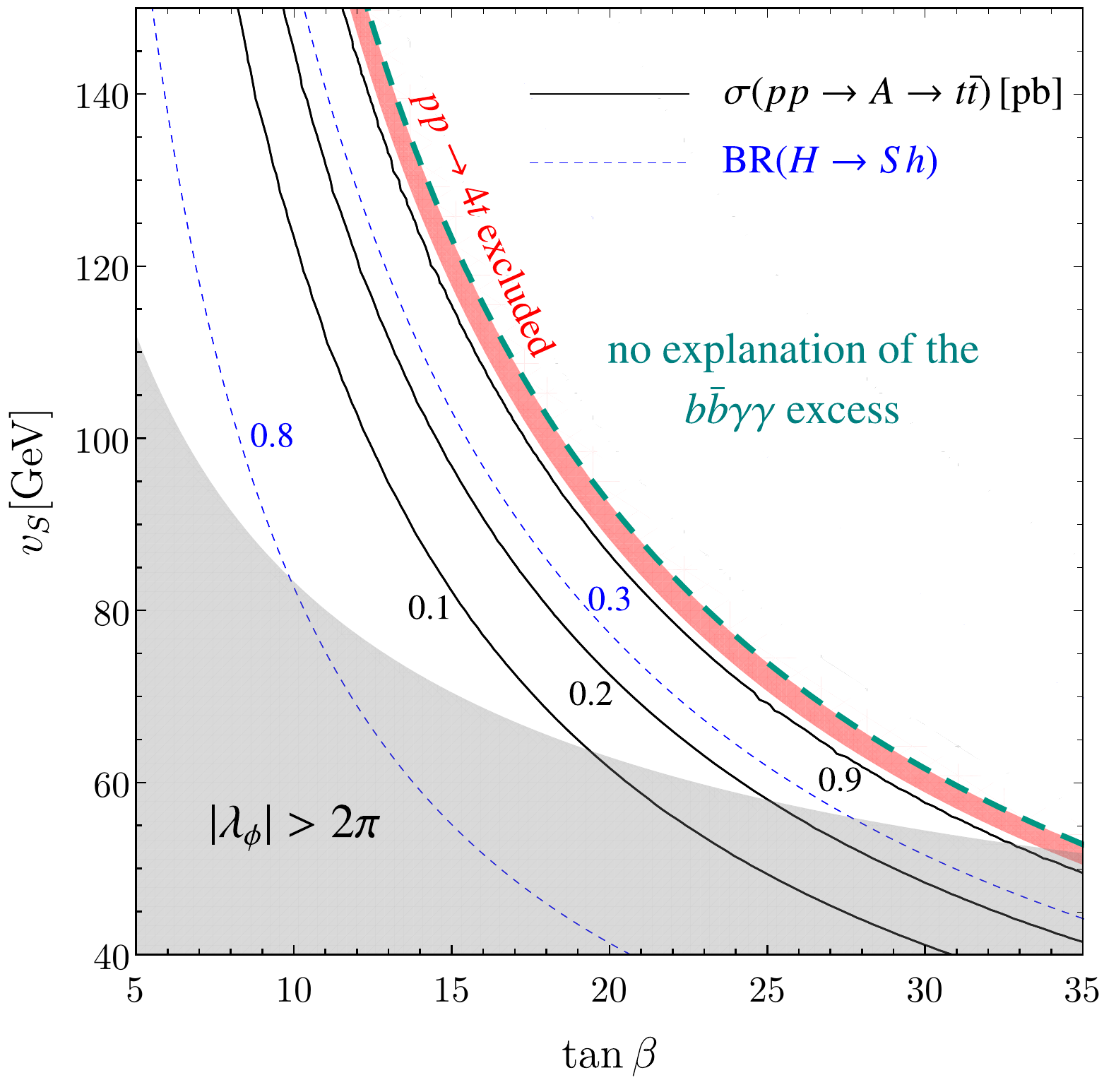}
    \caption{Left: Feynman diagram showing resonant asymmetric Higgs pair production. The discovery of this process, for which the CMS measurement constitutes a first hint, would be a smoking gun for the N2HMD-$U(1)$. Here, the black circle denotes the loop-induced effective coupling to gluons. However, note that the heavy top limit cannot be used because $m_t \ll m_H$ and we use the expression for a dynamical top quark in our numerical analysis. Right:
    Predictions for $\sigma(pp\to A\to t\bar t)$[pb] as a function of $\tan\beta$ and $v_S$ in the N2HMD-$U(1)$ for case (a), assuming that the CMS excess $b\bar b \gamma\gamma $ is explained. The grey region is excluded by the requirement of perturbative couplings, while the red region is excluded by the $t\bar t t \bar t$ search \cite{ATLAS:2022rws}, assuming $m_A\approx m_H$. Note that the $b\bar b \gamma\gamma $ excess cannot be explained in the top-right region of the green dashed line.}
    \label{fig:my_label}
\end{figure}

\subsection{Neutral scalars at the LHC ($\gamma\gamma$)}

These hints point towards the extension of the scalar sector of the SM by at least one $SU(2)_L$ doublet~\cite{Haisch:2017gql,Biekotter:2022abc}. In case, one aims at also addressing the di-Higgs excess (i.e.~$650\,{\rm GeV}\to b\bar b(90\,{\rm GeV})+\gamma\gamma(125\,{\rm GeV})$), the resonant pair production of the SM Higgs and a new scalar is required. This is, in fact, a natural signal of the N2HDM-$U(1)$~\cite{Banik:2023ecr} whose predictions are shown in Fig.~\ref{fig:my_label}.

\subsection{(di-)di-jet (jj(jj))}~Here two options come to mind~\cite{Crivellin:2022nms}: two scalar di-quarks or new massive gluons seem to be the most plausible candidates. Concerning the latter, a specific example is based on an $SU(3)_1\times SU(3)_2\times SU(3)_3$ gauge group, broken down to $SU(3)$ color via two bi-triplets.

\subsection{$q\bar q\!\to\! e^+e^-$}~As this analysis involves “non-resonant” electrons that do not originate from the on-shell production of a new particle, NP must be heavier than the energy scale of the LHC. This can be achieved with NP at the 10~TeV scale with order one coupling to first-generation quarks and electrons~\cite{Crivellin:2022rhw}. Therefore, $Z^\prime$~bosons~\cite{Crivellin:2021bkd} or LQs~\cite{Crivellin:2021egp} have the potential to explain the CMS measurement.

\subsection{Summary}

\begin{figure}
    \centering    
    \includegraphics[scale=0.6]{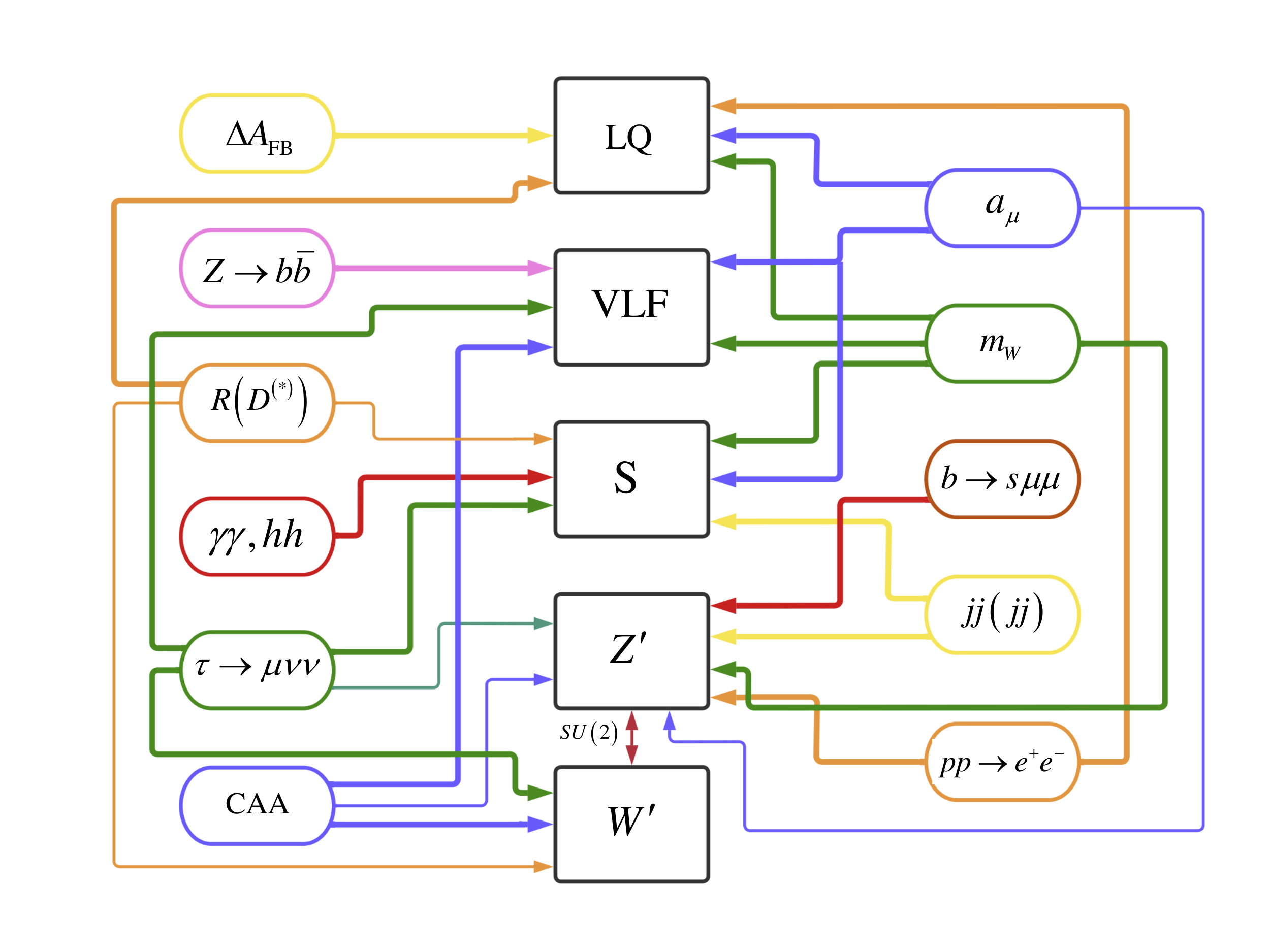}
    \caption{Summary of the anomalies together with the implications for extending the SM with new particles: Leptoquarks (LQ), vector-like fermions (VLF), electrically neutral scalars (S), neural gauge bosons ($Z^\prime$) and charged gauge bosons ($W^\prime$). Thick lines indicate that full explanations are possible while thin lines mean that only a partial one is  or that conflicts with other observables exist.}
    \label{summary}
\end{figure}

The anomalies, together with the extensions of the SM to which they point, are shown in Fig.~\ref{summary}.

\end{document}